\documentclass[reprint]{revtex4-2}
\usepackage{graphicx} 
\usepackage{dcolumn} 
\usepackage{bm} 
\usepackage{amsmath}
\usepackage{url}
\begin{document}

\title{Maximum likelihood estimation for reversible mechanistic network models}
\author{Jonathan Larson}
\author{Jukka-Pekka Onnela}
\affiliation{Department of Biostatistics, Harvard University, Boston, Massachusetts 02115}

\begin{abstract}
Mechanistic network models specify the mechanisms by which networks grow and change, allowing researchers to investigate complex systems using both simulation and analytical techniques. Unfortunately, it is difficult to write likelihoods for instances of graphs generated with mechanistic models, and thus it is near impossible to estimate the parameters using maximum likelihood estimation. In this paper, we propose treating node sequence in a growing network model as an additional parameter, or as a missing random variable, and maximizing over the resulting likelihood. We develop this framework in the context of a simple mechanistic network model, used to study gene duplication and divergence, and test a variety of algorithms for maximizing the likelihood in simulated graphs. We also run the best-performing algorithm on one human protein-protein interaction network and four non-human protein-protein interaction networks. Although we focus on a specific mechanistic network model, the proposed framework is more generally applicable to reversible models.
\end{abstract}

\maketitle

\section{Introduction}
\label{sec:intro}

The landscape of network models is dominated by two major types: statistical and mechanistic. Statistical models specify a likelihood for each instance of a graph, usually in terms of some sufficient statistic like the number of edges. For example, consider the $G(n,p)$ graph \cite{gilbert1959}. It has $n$ nodes, and each of the $\binom{n}{2}$ pairs of nodes (also called dyads) is connected by an edge independently and with probability $p$. If we consider $n$ to be known, $p$ to be unknown, and $x$ to be the number of edges in the observed graph, then conditional on $x$, the ``location'' of the edges (i.e., which dyads are connected) does not depend on $p$. Thus, $x$ is a sufficient statistic for $p$, and since $x$ is binomially distributed with parameters $\binom{n}{2}$ and $p$, we can write the likelihood for $x$ and carry out maximum likelihood (ML) estimation of $p$. In this simple case, the ML estimate of $p$ is given by $\hat{p} = x / \binom{n}{2}$.

Mechanistic network models are harder to characterize with a likelihood.
Each such model specifies a set of mechanisms
by which a network grows and changes over time.
Although these mechanisms may consist of relatively simple steps,
each step may erase the evidence of past steps. For example, consider the duplication-mutation-complementation (DMC)
model \cite{vazquezetal2003,vazquez2003}.
It was intended to model how a protein-protein interaction
network evolves.
In a protein-protein interaction network,
each node represents a protein in an organism 
and two proteins are connected if they interact.
In the DMC model,
the gene encoding a protein is erroneously duplicated with probability $q_m$,
and the duplicated gene produces an identical protein.
Over time, due to evolutionary pressure, the duplicate genes mutate separately,
leading to two new proteins that may interact with different sets of proteins. The original and duplicated proteins interact with probability $q_c$.

The DMC graph is similar to the $G(n,p)$ graph in that it is generated
through a series of Bernoulli random experiments.
It differs because the outcomes of those experiments cannot be determined
just by looking at the graph observed at the end.
The addition of each new node can drastically change
the relationships among previous nodes,
erasing the evidence of previous duplication, modification,
and complementation events.
We can write down a likelihood for any given graph,
but this becomes intractable when there are more than a handful of nodes.
This is typical of mechanistic network models,
which may be analyzed using
likelihood-free inference \cite{chen2019,chen2020}.

There is a growing literature on statistical inference and model selection methods for mechanistic network models. Different papers use different approaches for addressing the intractability of the likelihood: they compare summary statistics by considering a trade-off between their informativeness and computational cost \cite{raynal2023}; approximate the likelihood of a class of duplication-attachment models using an importance sampling scheme \cite{wiuf2006}; seek to identify a change point in network growth mechanisms using a likelihood-based framework that assumes the history of the network is known ~\cite{arnold2021}; make the likelihood tractable by modeling network formation through conditional multinomial logit models from discrete choice theory assuming that edge formation events are observed \cite{overgoor2019}; and by inferring the node history for a random tree growth model \cite{cantwell2021}. 

In this paper, we propose a method for conducting maximum likelihood
estimation of the parameters of growing mechanistic network models.
We use the DMC model as a test case, and thus our goal is to estimate $q_m$ and $q_c$ from a single observation of the network.

\section{Model}
\label{sec:model}
\subsection{DMC graph model}

The algorithm for generating a DMC graph with $n$ nodes is
as follows. We first begin with a seed graph. Throughout this paper, we will use a single node as the seed graph. Repeat the following three steps until the graph has $n$ nodes.  \textbf{Duplication:} (i) Select a node uniformly at random; this will be the ``anchor'' node. (ii) Add a new node. (iii) Connect the new node to the anchor node's neighbors. \textbf{Mutation:} (i)``Modify'' each of the anchor node's neighbors independently and with probability $q_m$. (ii) If a neighbor is modified, remove \textit{either} the neighbor-anchor node edge \textit{or} the neighbor-new node edge. Which edge is lost is determined by the flip of a fair coin. \textbf{Complementation:} (i) Connect the new node to the anchor node with probability $q_c$.

\begin{figure}
\centering
\includegraphics[width = 1\linewidth]{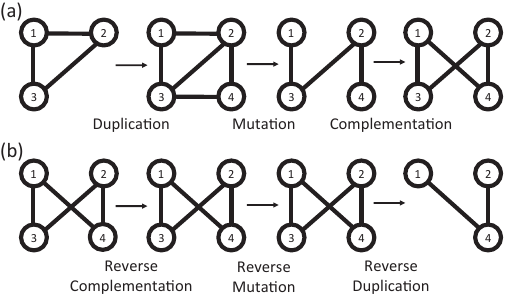}
\caption{\label{figure:schematic} (a) Schematic of the DMC model mechanisms, where 
node 1 is the anchor node and node 4 is the new node. 
(b) Schematic of the reverse DMC mechanisms,
where node 3 is incorrectly assumed to be the new node
and node 4 is incorrectly assumed to be the anchor node.}
\end{figure}

Figure \ref{figure:schematic}(a) contains a schematic of the DMC model mechanisms. In the duplication step, node 4 (the new node)
duplicates node 1 (the anchor node).
In the mutation step,
both of node 1's neighbors are modified,
with node 2 losing its edge with node 1
and node 3 losing its edge with node 4.
In the complementation step,
node 4 connects to node 1.

Middendorf \textit{et al.} \cite{middendorf2005} found the DMC model to explain
the observed \textit{D. melanogaster} protein-protein interaction network
better than six other candidate models.
The authors argue that, in the modification step,
it makes sense to remove either the neighbor-anchor node edge
or the neighbor-new node edge, but not both,
because each of these edges represents a function originally performed by
the anchor node.
If both disappeared, that would mean that both the gene coding for the
anchor node protein and the gene coding for the new node protein
mutated so much that neither could perform that function anymore.
Since the function was presumably necessary, 
at least one of these edges must be maintained.
The authors also note that the DMC model only allows for new edges
in the context of duplication and complementation,
meaning that mutations that result in brand new, advantageous
functions performed by the protein are rare.

\subsection{Graph deconstruction}

Consider if we knew not just the (final) topology of the graph
but also when each node entered the graph and which
existing node served as its anchor node.
Since each step of the DMC mechanism is reversible,
we could run the evolution of the graph backward,
determine the outcomes of the Bernoulli experiments,
and estimate $q_m$ and $q_c$.

Let $G_n$ denote the observed graph of $n$ nodes. Let $\theta_n = (\alpha_n,\beta_n)$, where $\alpha_n$ is an $n$-vector containing the sequence in which the nodes entered the graph, and $\beta_n$ is an $n$-vector containing the sequence of anchor nodes. Thus the $i$th element of $\beta_n$ was the anchor node for the $i$th element of $\alpha_n$. (The first node, our seed graph, has no anchor node; if the nodes of the graph are labelled from 1 to $n$, we set the anchor node of the first node to be 0.)

Repeat the following three steps for $i = n,n-1,\ldots,2$.
\textbf{Reverse Complementation:}
(i) Set
$W_i(G_n,\theta_n) = 1$ if the new node
(the $i$th node in $\alpha_n$)
and its anchor node
(the $i$th node in $\beta_n$)
are connected,
and $W_i(G_n,\theta_n) = 0$ otherwise.
(ii) If the new node and the anchor node are connected,
remove that edge.
\textbf{Reverse Mutation:}
(i) Set $X_i(G_n,\theta_n)$ equal to the cardinality
of the intersection of the
neighbors of the anchor node and the new node.
This is the number of unmodified neighbors of the anchor node.
(ii) Set $Y_i(G_n,\theta_n)$ equal to the cardinality of
the union of the neighbors of the anchor node and the new node.
This is the original degree of the anchor node,
before it was modified.
(iii) Connect the anchor node to all the neighbors of the new node,
and connect the new node to all the neighbors of the anchor node.
\textbf{Reverse Duplication:}
(i) Remove the new node.

Here $W_i$, $X_i$, and $Y_i$ are written as functions of $G_n$ and $\theta_n$
because their values are completely determined by $G_n$ and $\theta_n$. Typically the true insertion order of nodes is however not known. Figure \ref{figure:schematic}(b) contains a schematic of this process
for $\theta_n'$, which is not equal to the true value $\theta_n$.
Here, node 3 is assumed to be the new node and node 4 is assumed
to be the anchor node.
In the reverse complementation step,
nodes 3 and 4 are not connected by an edge,
so $W_i(G_n,\theta_n') = 0$.
In the reverse mutation step,
nodes 3 and 4 have exactly the same neighbors, nodes 1 and 2,
so $X_i(G_n,\theta_n') = Y_i(G_n,\theta_n') = 2$.
Finally, in the reverse duplication step,
node 3 is assumed to be the new node and is thus removed.
Note that for the true value of $\theta_n$,
$W_i(G_n,\theta_n) = 1$,
$X_i(G_n,\theta_n) = 0$,
and $Y_i(G_n,\theta_n) = 2$.

\subsection{Inference}
If we think of $\theta_n$ as a parameter like $q_m$ and $q_c$,
and define
\begin{align}
W(G_n,\theta_n) &= \sum_{i=2}^n W_i(G_n,\theta_n) \\
X(G_n,\theta_n) &= \sum_{i=2}^n X_i(G_n,\theta_n) \\
Y(G_n,\theta_n) &= \sum_{i=2}^n Y_i(G_n,\theta_n) \text{,}
\end{align}
then we can write the likelihood (likelihood function) as
\begin{align}
\mathcal{L}(q_m,q_c,\theta_n) &= q_m^{Y(G_n,\theta_n) - X(G_n,\theta_n)} (1-q_m)^{X(G_n,\theta_n)} \nonumber \\
&\times q_c^{W(G_n,\theta_n)} (1-q_c)^{n - 1 - W(G_n,\theta_n)} \text{.}
\end{align}

The likelihood is defined as the joint probability of the observed data as a function of the model parameters. In practice, it is common to omit constants that do not depend on the model parameters, in which case the likelihood is only defined up to a multiplicative constant of proportionality. For example, the binomial distribution with parameters $n$ and $p$ specifies the probability distribution of the number of successes in a sequence of $n$ independent experiments; the binomial likelihood function $\mathcal{L}(p|n,y) = \binom{n}{y} p^y (1-p)^{n-y}$ without the multiplicative constant would be written as $\mathcal{L}(p) = p^y (1-p)^{n-y}$. In the DMC model, the likelihood is a product of two independent binomial likelihoods, one for the mutation process and one for the complementation process, which is how we arrive at the above expression.

The most common approach to parameter estimation is to select parameter values that assign the highest probability to the observed data. This is known as maximum likelihood estimation (MLE). In our case, we can maximize this likelihood with
\begin{align}
\hat{q}_m &= 1 - \frac{X(G_n,\theta_n)}{Y(G_n,\theta_n)} \\
\hat{q}_c &= \frac{W(G_n,\theta_n)}{n - 1} \\
\hat{\theta}_n &= \arg\max_{\theta_n} \mathcal{L}(\hat{q}_m,\hat{q}_c,\theta_n)  \text{.}
\end{align}

Right away, there are some caveats worth mentioning.
First, different values of $\theta_n$ may yield the same
likelihood for a given $G_n$,
so $\theta_n$ may not be identifiable.
Also, $\theta_n$ resides in a space of dimension $n!(n-1)!$,
which increases rapidly with $n$.
However,
since the duplication step renders the new node identical to the anchor node,
it does not matter which is labeled new and which is labeled anchor.
Thus the parameter $\theta_n$ really only needs to encode the sequence
of pairs used to deconstruct the graph,
of which there are
\begin{equation}
\prod_{i=2}^{n} \binom{i}{2} = \frac{n!(n-1)!}{2^{n-1}}
\end{equation}
possible values.
Even though this is less than $n!(n-1)!$,
it grows rapidly with $n$.
For the rest of this paper,
for clarity, we will consider $\theta_n = (\alpha_n,\beta_n)$
as having $n!(n-1)!$ possible values.

The idea of estimating the age of each node
(that is, when each node entered the graph) is not new.
Navlakha and Kingsford \cite{navlakha2011} coined the term
``network archaeology'' for this endeavor,
and proposed the following algorithm for estimating $\alpha_n$:
\begin{enumerate}
\item Select initial values of $q_m$ and $q_c$.
\item \label{stepstart} For each pair of nodes $(u,v)$ in the graph,
do the following:
\begin{enumerate}
\item Let $W(u,v) = 1$ if $u$ and $v$ are connected by an edge
and $0$ if they are not.
\item Let $X(u,v)$ equal the cardinality of the intersection of the
neighbors of $u$ and $v$.
\item Let $Y(u,v)$ equal the cardinality of the union of the
neighbors of $u$ and $v$.
\item Using the initial values of $q_m$ and $q_c$, compute
\begin{align}
\mathcal{L}(u,v) &= q_c^{W(u,v)} (1-q_c)^{1 - W(u,v)} \nonumber \\
& \times q_m^{Y(u,v) - X(u,v)} (1-q_m)^{X(u,v)} \text{.}
\end{align}
\end{enumerate}
\item \label{stepend} Select the pair of nodes $(u,v)$
that maximizes $\mathcal{L}(u,v)$
and reverse the DMC mechanism using those nodes.
\item Repeat steps \ref{stepstart} through \ref{stepend} until the graph
has one node.
\end{enumerate}
Navlakha and Kingsford \cite{navlakha2011} sought
to estimate $\alpha_n$, not $(q_m,q_c)$,
although they did suggest using likelihoods to select optimal
values of $q_m$ and $q_c$.

Arguably, $\theta_n$ is not a parameter but a missing random variable
that does not depend on $q_m$ or $q_c$.
This framework lends itself to expectation-maximization (EM)
\cite{dempster1977},
which maximizes an intractable likelihood for incomplete data
(in this case, the graph)
by using a tractable likelihood for complete data
(in this case, the graph and the sequence of new and anchor nodes).
EM also avoids the problems outlined in \cite{meng2009}.
Let $Z_n = (U_n,V_n)$,
where $U_n$ is an $n$-vector containing the sequence in which the nodes
entered the graph,
and $V_n$ is an $n$-vector containing the sequence of anchor nodes.
Thus the $i$th element of $V_n$ was the anchor node
for the $i$th element of $U_n$.
(The first node, the seed graph, has no anchor node, and we simply set its anchor node to 0.) Then the probability of observing $G_n$ and $Z_n$ is

\begin{widetext}

\begin{equation}
f(G_n,Z_n;q_m,q_c) = \frac{1}{n!(n-1)!} q_m^{Y(G_n,Z_n) - X(G_n,Z_n)} 
 (1-q_m)^{X(G_n,Z_n)} q_c^{W(G_n,Z_n)} (1-q_c)^{n - 1 - W(G_n,Z_n)} \text{.}
\end{equation}

Then the EM $Q$-function, which is the expected value of the log-likelihood function with respect to the conditional distribution of the unobserved node sequence given the observed graph $G_n$ and the current estimates of parameters $q_m'$ and $q_c'$, is given by
\begin{align}
Q(q_m'',q_c''|q_m',q_c') = \sum_z f(G_n,z;q_m',q_c') \log f(G_n,z;q_m'',q_c'') \text{,}
\end{align}
where the sum is across all $n!(n-1)!$
possible sequences of new and anchor nodes.
Given initial values $q_m'$ and $q_c'$,
the E-step consists of calculating $Q(q_m'',q_c''|q_m',q_c')$
and the M-step consists of finding

\begin{equation}
\arg\max_{(q_m'',q_c'')} Q(q_m'',q_c''|q_m',q_c') = \left(1 - \frac{\sum_z f(G_n,z;q_m',q_c') X(G_n,z)}{\sum_z f(G_n,z;q_m',q_c') Y(G_n,z)},\frac{\sum_z f(G_n,z;q_m',q_c') W(G_n,z)}{n - 1}\right) \text{.}    
\end{equation}
\end{widetext}

These maxima serve as initial values $q_m'$ and $q_c'$
in the next round of EM,
and the process repeats until the $Q$-function converges.

Regardless of whether we think of the sequence of new and anchor nodes
as a parameter or an unknown random variable,
it is not what we want to know.
We are interested in estimating $q_m$ and $q_c$.
Thus, it might behoove us to average over all possible values of $\theta_n$.
Taking our cue from the EM algorithm,
we could use the estimates

\begin{widetext}
\begin{equation}
(\hat{q}_m, \hat{q}_c) = 
\left(1 - \frac{\sum_{\theta_n} \mathcal{L}(\hat{q}_m,\hat{q}_c,\theta_n) X(G_n,\theta_n)}{\sum_{\theta_n} \mathcal{L}(\hat{q}_m,\hat{q}_c,\theta_n) Y(G_n,\theta_n)},\frac{\sum_{\theta_n} \mathcal{L}(\hat{q}_m,\hat{q}_c,\theta_n) W(G_n,\theta_n)}{n - 1}\right) \text{.}
\end{equation}
\end{widetext}
Of course, this all supposes that we can exhaustively search all
possible values of $\theta_n$ (or $Z_n$).
But this is infeasible for graphs with more than a handful of nodes.
Thus, we need to find algorithms that can find reasonably
good values of $\theta_n$ in a reasonable amount of time.
We tested a variety of algorithms on DMC graphs generated
with a variety of values of $q_m$ and $q_c$.
We also tested the best-performing algorithm on an empirical
protein-protein interaction network as described below.

\section{Methods}
\subsection{Model-generated graphs}

We generated small DMC graphs with 7, 100, and 200 nodes.
For each of these three graph sizes,
we generated 100 graphs,
using every possible pair of $(q_m,q_c)$ where $q_m,q_c \in \left\{1/11, 2/11 ,\dotsc,10/11\right\}$. We intentionally avoided the boundaries
$q_m,q_c \in \{0,1\}$ in order for the graphs
to be random.
For example,
since we used a single node as the seed graph,
setting $q_c = 0$ would have generated a graph with zero edges,
and setting $(q_m,q_c) = (0,1)$ would have generated a complete graph.
We then ran each of the following deconstruction algorithms on each graph:
\begin{enumerate}
\item True $\theta_n$:
This algorithm uses the true value of $\theta_n$.
\item True New, Random Anchor:
This algorithm uses the true sequence of new nodes
but selects anchor nodes uniformly at random.
\item NK, True Initial:
This is the algorithm from \cite{navlakha2011},
described in the Introduction,
using the true values of $q_m$ and $q_c$ as initial values.
\item Exhaustive:
This is an exhaustive search of all possible values of $\theta_n$.
\item NK:
This is the algorithm from \cite{navlakha2011},
described in the Introduction,
using all possible pairs of $(q_m,q_c)$ where
$q_m,q_c \in \left\{1/5, 2/5, 3/5, 4/5\right\}$ as initial values.
The estimate $(\hat{q}_m,\hat{q}_c)$ that yields the highest likelihood
is used as the center of a new four-by-four grid of values,
this time with gaps of $\frac{1}{5^2}$ instead of $\frac{1}{5}$.
This process of using finer and finer grids of values
is repeated until the likelihood stops increasing.
\item NK+1:
This algorithm is the same as NK,
except it uses one additional grid of values after the likelihood
stops increasing.
Due to time constraints,
we only ran this algorithm on graphs with 7 and 100 nodes.
\item Minimize $Y(u,v)$:
Given the current state of the graph,
this algorithm selects the pair of nodes $(u,v)$ with the lowest value
of $Y(u,v)$, the cardinality of the union of the
neighbors of $u$ and $v$.
(If multiple pairs share the lowest value of $Y(u,v)$,
one of these pairs is selected uniformly at random.)
It then reverses the DMC mechanism using that pair of nodes
and repeats.
It is based on the fact that
\begin{equation}
\frac{\partial \log \mathcal{L}(u,v)}{\partial Y(u,v)} =  \log q_m < 0
\end{equation}
and thus $\mathcal{L}(u,v)$ is maximized by minimizing $Y(u,v)$.
(The sign of the partial derivative of $\log \mathcal{L}(u,v)$
taken with respect to $W(u,v)$ depends on the value of $q_c$;
the sign of the partial derivative of $\log \mathcal{L}(u,v)$
taken with respect to $X(u,v)$ depends on the value of $q_m$.
Thus neither of these seemed to be a good criterion for selecting
the next pair of nodes to deconstruct the graph.)
\item Minimize $Y(u,v)$, then NK:
This algorithm runs the Minimize $Y(u,v)$ algorithm and then uses
the resulting estimates $\hat{q}_m$ and $\hat{q}_c$ as initial values
for the NK algorithm.
The resulting estimates are fed back into NK as initial values,
and this process is repeated until the likelihood stops increasing.
\item 1 Random:
Given the current state of the graph,
this algorithm selects a pair of nodes uniformly at random
and runs the DMC mechanism backwards.
\item 100 Random:
This algorithm runs the 1 Random algorithm 100 times.
\end{enumerate}

Some of the algorithms
(Exhaustive; NK; NK+1; Minimize $Y(u,v)$, then NK; and 100 Random)
yield multiple values of $\theta_n$.
For each of these,
we selected the value of $\theta_n$ that maximized
$\mathcal{L}(\hat{q}_m,\hat{q}_c,\theta_n)$.
In order to take advantage of the data on several values of $\theta_n$,
we also conducted Expectation-Maximization (EM) as described
in the Introduction,
except we summed over the investigated values of $\theta_n$ (or $z$)
and not all possible values of $\theta_n$ (or $z$).
We used the maximum-likelihood values of $\hat{q}_m$ and $\hat{q}_c$
as initial values.
Similarly, we averaged across the investigated values of $\theta_n$
as described in the Introduction.

We computed 95\% confidence intervals for each algorithm's estimates
of $q_m$ and $q_c$ as follows:
\begin{equation}
\hat{q}_m \pm 1.96\sqrt{\frac{\hat{q}_m \left(1 - \hat{q}_m\right)}{Y(G_n,\theta_n)}} \nonumber,
\hat{q}_c \pm 1.96\sqrt{\frac{\hat{q}_c \left(1 - \hat{q}_c\right)}{n - 1}} \text{.}
\end{equation}

For the algorithms that yielded multiple values of $\theta_n$,
we selected the maximum-likelihood value
when computing the confidence intervals.

In addition to obtaining estimates of $q_m$ and $q_c$ from each algorithm,
we also calculated Kendall's $\tau$ \cite{kendall1938}
to compare the true sequence of new nodes to the estimated sequence
of new nodes.
Kendall's $\tau$ considers each of the $\binom{n}{2}$ pairs of nodes
in the graph.
If the relative ordering of the nodes in the pair is correct
in the estimated sequence of new nodes,
this pair is considered concordant.
If the relative ordering of the nodes in the pair is incorrect
in the estimated sequence of new nodes,
this pair is considered discordant.
Kendall's $\tau$ is equal to the number of concordant pairs
less the number of discordant pairs,
divided by the total number of pairs.
If the estimated sequence of new nodes is in the exact right order,
$\tau = 1$.
If the estimated sequence of new nodes is in reverse order,
$\tau = -1$.

Since the duplication step of the DMC model renders the new node and
anchor node indistinguishable,
we computed two different values.
The Strict Kendall's $\tau$ assumes that each algorithm cannot distinguish
between the anchor node and new node when deconstructing the graph,
and thus must select one of them at random to remove.
The Lenient Kendall's $\tau$ assumes that each algorithm can tell
which node in a pair was the anchor node and which was the new node,
and thus removes the new node when deconstructing the graph.

We generated four additional 100-node DMC graphs
with parameters
\[
(q_m,q_c) \in \left\{\left(\frac{1}{3},\frac{1}{3}\right), \left(\frac{1}{3},\frac{2}{3}\right), \left(\frac{2}{3},\frac{1}{3}\right), \left(\frac{2}{3},\frac{2}{3}\right)\right\} \text{.}
\]
For each, we ran the following algorithms:
True $\theta_n$; True New, Random Anchor; NK, True Initial;
NK; Minimize $Y(u,v)$; Minimize $Y(u,v)$, then NK; and 100 Random.
For each graph, we plotted the log-likelihood as a function of the
estimates of $q_m$ and $q_c$.

\subsection{Empirical graphs}

To test our approach on an empirical network, we  obtained the human protein-protein interaction network described in \cite{luck2020} from \cite{interactome2021}. This is the largest and most recent protein-protein interactome obtained from humans. The obtained Human Reference Interactome (HuRI) graph contained 8,272 nodes and 52,548 edges. We deleted 480 self-loops because the DMC model does not allow for them, yielding 52,068 edges. Then we sampled $p = 0.05,0.10,\dotsc,0.90,0.95$ of the nodes from HuRI uniformly at random and ran the Minimize $Y(u,v)$ algorithm on each induced subgraph. We also ran the Minimize $Y(u,v)$ algorithm on the complete HuRI graph.

We also obtained protein-protein interaction networks for the following
organisms from the STRING database \cite{szklarczyk2018}:
\textit{C. elegans},
\textit{D. melanogaster},
\textit{E. coli}, and
\textit{S. cerevisiae}.
For each of these four organisms, we downloaded the dataset labeled
``protein network data (scored links between proteins)'' which
contained only physical links.
We removed self loops
and links with combined confidence scores less than or equal to 700.
Then we ran Minimize $Y(u,v)$ on the resulting graphs.

For the HuRI graph and the four STRING graphs,
we sampled ten percent of the nodes uniformly at randomed
and ran Minimize $Y(u,v)$ on the induced subgraph.
Using the estimates of $q_m$ and $q_c$ from each subgraph,
we generated a DMC graph with the same number of nodes
and ran Minimize $Y(u,v)$ on it as well.

We conducted all simulations on the O2 High Performance Compute Cluster,
supported by the Research Computing Group, at Harvard Medical School.
We conducted all analyses and simulations in R Version 3.6.1 \cite{r2019},
except for the Minimize $Y(u,v)$ algorithm that we ran
on the HuRI and STRING graphs.
We conducted those analyses in Python Version 3.7.4.

\section{Results}
\subsection{Model-generated graphs}

The root mean squared errors (RMSEs)
for $q_m$ and $q_c$ for model-generated graphs are in Table \ref{table:max}.
Minimize $Y(u,v)$ has the best performance
(among algorithms that do not require prior knowledge of $\theta_n$,
$q_m$, or $q_c$)
for both parameters on graphs of 200 nodes;
and for $q_c$ on graphs of 100 and 7 nodes.
Its performance for $q_m$ on graphs of 100 nodes is second only
to Minimize $Y(u,v)$, then NK.

\begin{table}
\begin{ruledtabular}
\begin{tabular}{ccccccc}
\multicolumn{1}{c}{} & \multicolumn{6}{c}{\textbf{RMSE}} \\
\multicolumn{1}{c}{\textbf{Method}} & \multicolumn{2}{c}{\textbf{7 Nodes}} & \multicolumn{2}{c}{\textbf{100 Nodes}} & \multicolumn{2}{c}{\textbf{200 Nodes}} \\
 & $\bm{q_m}$ & $\bm{q_c}$ & $\bm{q_m}$ & $\bm{q_c}$ & $\bm{q_m}$ & $\bm{q_c}$ \\
\hline
1 & 0.178 & 0.177 & 0.034 & 0.037 & 0.024 & 0.028 \\
2 & 0.332 & 0.229 & 0.360 & 0.318 & 0.373 & 0.357 \\
3 & 0.209 & 0.168 & 0.067 & 0.100 & 0.071 & 0.103 \\
4 & 0.431 & 0.262 & N/A & N/A & N/A & N/A \\
5 & 0.430 & 0.262 & 0.138 & 0.136 & N/A & N/A \\
6 & 0.428 & 0.262 & 0.137 & 0.137 & N/A & N/A \\
7 & 0.327 & 0.205 & 0.095 & 0.109 & 0.074 & 0.107 \\
8 & 0.311 & 0.251 & 0.126 & 0.144 & 0.069 & 0.129 \\
9 & 0.320 & 0.253 & 0.360 & 0.321 & 0.376 & 0.350 \\
10 & 0.380 & 0.270 & 0.345 & 0.296 & 0.366 & 0.347 \\
\end{tabular}
\end{ruledtabular}
\caption{\label{table:max} Root mean squared error (RMSE)
for $q_m$ and $q_c$ and for each graph size.
For algorithms that yielded multiple values of $\theta_n$
(Exhaustive; NK; NK+1; Minimize $Y(u,v)$, then NK; and 100 Random),
the results shown here are for the value of $\theta_n$ that yielded
the highest likelihood.
``N/A'' indicates that this algorithm was not run on graphs of this size
because of time constraints.
The worst possible RMSE, achieved if $\hat{q}_m = 1$
whenever $q_m \le 0.5$ and $\hat{q}_m = 0$
whenever $q_m > 0.5$, is 0.739.}
\end{table}

It is worth noting that the Exhaustive algorithm has the worst performance
for $q_m$ on graphs of 7 nodes.
Supplemental Table 1 may explain why.
The likelihood is maximized when $\hat{q}_m$ and $\hat{q}_c$ are 0 or 1.
If there exists a value of $\theta_n$ such that
$\hat{q}_m$ and/or $\hat{q}_c$ is 0 or 1,
the Exhaustive algorithm will find it.
As Supplemental Table 1 shows,
the Exhaustive algorithm estimates $\hat{q}_m$ to be 0 or 1 more
than any other algorithm,
and it estimates $\hat{q}_c$ to be 0 or 1 more than
any other algorithm except 100 Random.
The Exhaustive algorithm also has the greatest bias
for the log-likelihood,
meaning it is overestimating the likelihood
more than any other algorithm. The table also contains the number of times
each algorithm could not estimate $q_m$,
which happens when $Y(G_n,\theta_n) = 0$.
Since the Minimize $Y(u,v)$ algorithm minimizes $Y(u,v)$,
it could not estimate $q_m$ for the greatest number of graphs.

Table \ref{table:em} contains the RMSE for the expectation-maximization (EM)
and averaging results.
For graphs of 100 and 200 nodes, performance does not improve,
or it barely improves, when using EM or averaging instead of maximum likelihood.
For graphs of 7 nodes,
performance can be improved by using EM or averaging
instead of maximum likelihood.

\begin{table}
\begin{ruledtabular}
\begin{tabular}{lrrrrrr}
\multicolumn{1}{c}{} & \multicolumn{6}{c}{\textbf{RMSE}} \\
\multicolumn{1}{c}{\textbf{Method}} & \multicolumn{2}{c}{\textbf{7 Nodes}} & \multicolumn{2}{c}{\textbf{100 Nodes}} & \multicolumn{2}{c}{\textbf{200 Nodes}} \\
 & $\bm{q_m}$ & $\bm{q_c}$ & $\bm{q_m}$ & $\bm{q_c}$ & $\bm{q_m}$ & $\bm{q_c}$ \\
\hline
Exhaustive Max & 0.431 & 0.262 & N/A & N/A & N/A & N/A \\
Exhaustive EM & 0.438 & 0.271 & N/A & N/A & N/A & N/A \\
Exhaustive Ave & 0.364 & 0.256 & N/A & N/A & N/A & N/A \\
\hline
NK Max & 0.430 & 0.262 & 0.138 & 0.136 & N/A & N/A \\
NK EM & 0.432 & 0.262 & 0.137 & 0.136 & N/A & N/A \\
NK Ave & 0.353 & 0.252 & 0.134 & 0.135 & N/A & N/A \\
\hline
NK + 1 Max & 0.428 & 0.262 & 0.137 & 0.137 & N/A & N/A \\
NK + 1 EM & 0.429 & 0.263 & 0.136 & 0.138 & N/A & N/A  \\
NK + 1 Ave & 0.362 & 0.253 & 0.128 & 0.135 & N/A & N/A  \\
\hline
Min $Y(u,v)$, NK Max & 0.311 & 0.251 & 0.126 & 0.144 & 0.069 & 0.129 \\
Min $Y(u,v)$, NK EM & 0.312 & 0.250 & 0.126 & 0.144 & 0.069 & 0.129 \\
Min $Y(u,v)$, NK Ave & 0.310 & 0.250 & 0.125 & 0.142 & 0.068 & 0.128 \\
\hline
100 rnd sqs Max & 0.380 & 0.270 & 0.345 & 0.296 & 0.366 & 0.347 \\
100 rnd sqs EM & 0.393 & 0.267 & 0.346 & 0.297 & 0.366 & 0.347 \\
100 rnd sqs Ave & 0.363 & 0.248 & 0.346 & 0.297 & 0.366 & 0.347 \\

\end{tabular}
\end{ruledtabular}
\caption{\label{table:em} Root mean squared error (RMSE)
for $q_m$ and $q_c$ and for each graph size,
only for algorithms that yielded multiple values of $\theta_n$
(Exhaustive; NK; NK+1; Minimize $Y(u,v)$, then NK; and 100 Random).
``N/A'' indicates that this algorithm was not run on graphs of this size
because of time constraints. Abbreviations: rnd = random, sqs = sequences.
}
\end{table}

Observed coverage for the confidence intervals is in Table \ref{table:coverage}.
For every algorithm except True $\theta_n$,
coverage is low and gets lower as the number of nodes in the graph increases.

\begin{table}
\begin{ruledtabular}
\begin{tabular}{ccccccc}
\multicolumn{1}{c}{} & \multicolumn{6}{c}{\textbf{Coverage}} \\
\multicolumn{1}{c}{\textbf{Method}} & \multicolumn{2}{c}{\textbf{7 Nodes}} & \multicolumn{2}{c}{\textbf{100 Nodes}} & \multicolumn{2}{c}{\textbf{200 Nodes}} \\
 & $\bm{q_m}$ & $\bm{q_c}$ & $\bm{q_m}$ & $\bm{q_c}$ & $\bm{q_m}$ & $\bm{q_c}$ \\
\hline
1 & 0.68 & 0.73 & 0.96 & 0.96 & 0.97 & 0.98 \\
2 & 0.55 & 0.79 & 0.06 & 0.24 & 0.05 & 0.17 \\
3 & 0.55 & 0.72 & 0.60 & 0.59 & 0.36 & 0.48 \\
4 & 0.28 & 0.60 & N/A & N/A & N/A & N/A \\
5 & 0.30 & 0.60 & 0.53 & 0.51 & N/A & N/A \\
6 & 0.30 & 0.60 & 0.54 & 0.48 & N/A & N/A \\
7 & 0.70 & 0.73 & 0.40 & 0.54 & 0.34 & 0.39 \\
8 & 0.49 & 0.62 & 0.57 & 0.44 & 0.38 & 0.41 \\
9 & 0.60 & 0.75 & 0.05 & 0.23 & 0.04 & 0.18  \\
10 & 0.40 & 0.58 & 0.07 & 0.24 & 0.06 & 0.21 \\

\end{tabular}
\end{ruledtabular}
\caption{\label{table:coverage} Observed coverage for nominal 
95\% confidence intervals.
``N/A'' indicates that this algorithm was not run on graphs of this size
because of time constraints.}
\end{table}

Average values of Kendall's $\tau$ are in Table \ref{table:tau}.
The Strict version is near zero for any algorithm that doesn't
require prior knowledge of the true value of $\theta_n$.
When it comes to the Lenient version,
100 Random is the best-performing ``naive'' algorithm
for graphs with 7 and 100 nodes;
its performance is second only to 1 Random
for graphs with 200 nodes. 
Running times for the algorithms are in Supplemental Table 2.

\begin{table}
\begin{ruledtabular}
\begin{tabular}{ccccccc}
\multicolumn{1}{c}{} & \multicolumn{6}{c}{$\bm{\bar{\tau}}$} \\
\multicolumn{1}{c}{\textbf{Method}} & \multicolumn{2}{c}{\textbf{7 Nodes}} & \multicolumn{2}{c}{\textbf{100 Nodes}} & \multicolumn{2}{c}{\textbf{200 Nodes}} \\
 & \textbf{Strict} & \textbf{Lenient} & \textbf{Strict} & \textbf{Lenient} & \textbf{Strict} & \textbf{Lenient} \\
\hline
1 & 0.209 & 1.000 & 0.330 & 1.000 & 0.320 & 1.000 \\
2 & 0.350 & 1.000 & 0.331 & 1.000 & 0.339 & 1.000 \\
3 & -0.025 & 0.533 & 0.039 & 0.240 & 0.056 & 0.248 \\
4 & 0.006 & 0.556 & N/A & N/A & N/A & N/A \\
5 & -0.011 & 0.536 & 0.044 & 0.251 & N/A & N/A \\
6 & 0.058 & 0.567 & 0.044 & 0.255 & N/A & N/A \\
7 & -0.025 & 0.530 & 0.029 & 0.183 & 0.049 & 0.187 \\
8 & -0.025 & 0.526 & 0.030 & 0.218 & 0.055 & 0.223 \\
9 & -0.034 & 0.518 & 0.014 & 0.346 & 0.010 & 0.345 \\
10 & 0.040 & 0.564 & 0.006 & 0.364 & 0.000 & 0.349 \\
\end{tabular}
\end{ruledtabular}
\caption{\label{table:tau} Kendall's $\tau$ averaged across all graphs
of a given size.
The Lenient Kendall's $\tau$ is calculated as though each algorithm
always removes the older node in a given pair when deconstructing a graph.
The Strict Kendall's $\tau$ is calculated as though each algorithm
must choose which node in a given pair to remove at random when
deconstructing a graph.
``N/A'' indicates that this algorithm was not run on graphs of this size
because of time constraints.}
\end{table}

\begin{figure}
\centering
\includegraphics[width = 1\linewidth]{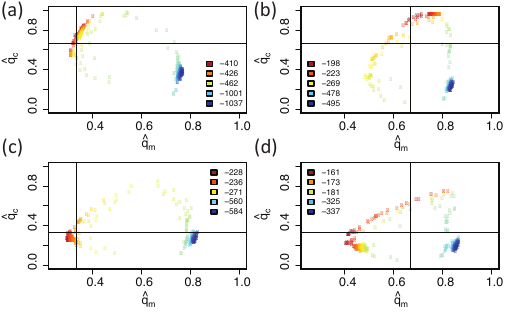}
\caption{\label{figure:nll} Log-likelihood as a function of the estimates
of $q_m$ and $q_c$.
The vertical and horizontal lines denote the true values of $q_m$ and $q_c$,
respectively, used to generate each graph.
The legend for each plot displays the quartiles of the log-likelihoods
and the corresponding colors.}
\end{figure}

\begin{figure}
\centering
\includegraphics[width = 1\linewidth]{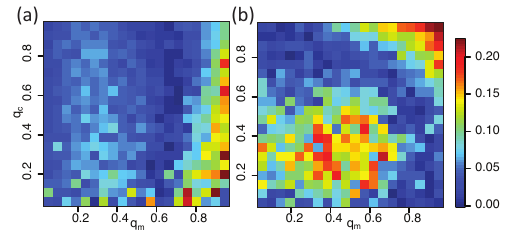}
\caption{\label{figure:error} Absolute error for the Minimize $Y(u,v)$
algorithm as a function of the true
values of $q_m$ and $q_c$ for 400-node graphs for (a) $\left|\hat{q}_m - q_m\right|$
and (b) $\left|\hat{q}_c - q_c\right|$.}
\end{figure}

Figure \ref{figure:nll} displays log-likelihood as a
function of $\hat{q}_m$ and $\hat{q}_c$
for four additional DMC graphs.
Each graph had 100 nodes
and $q_m,q_c\in\left\{\frac{1}{3},\frac{2}{3}\right\}$.
The log-likelihoods attained for each graph vary,
but the estimates appear in a characteristic oval pattern.
The algorithms appear to perform the worst on the graph with true
$(q_m,q_c) = \left(\frac{2}{3},\frac{1}{3}\right)$,
even though it has the highest log-likelihoods.
However, Figure \ref{figure:error} seems to indicate that,
when the Minimize $Y(u,v)$ algorithm is used on 400-node graphs,
the worst error for $\hat{q}_m$ should occur when $q_m$ is high
and the worst error for $\hat{q}_c$ should occur when
$q_m$ and $q_c$ are both below 0.6.

\subsection{Empirical graphs}

\begin{figure}
\centering
\includegraphics[width = 1\linewidth]{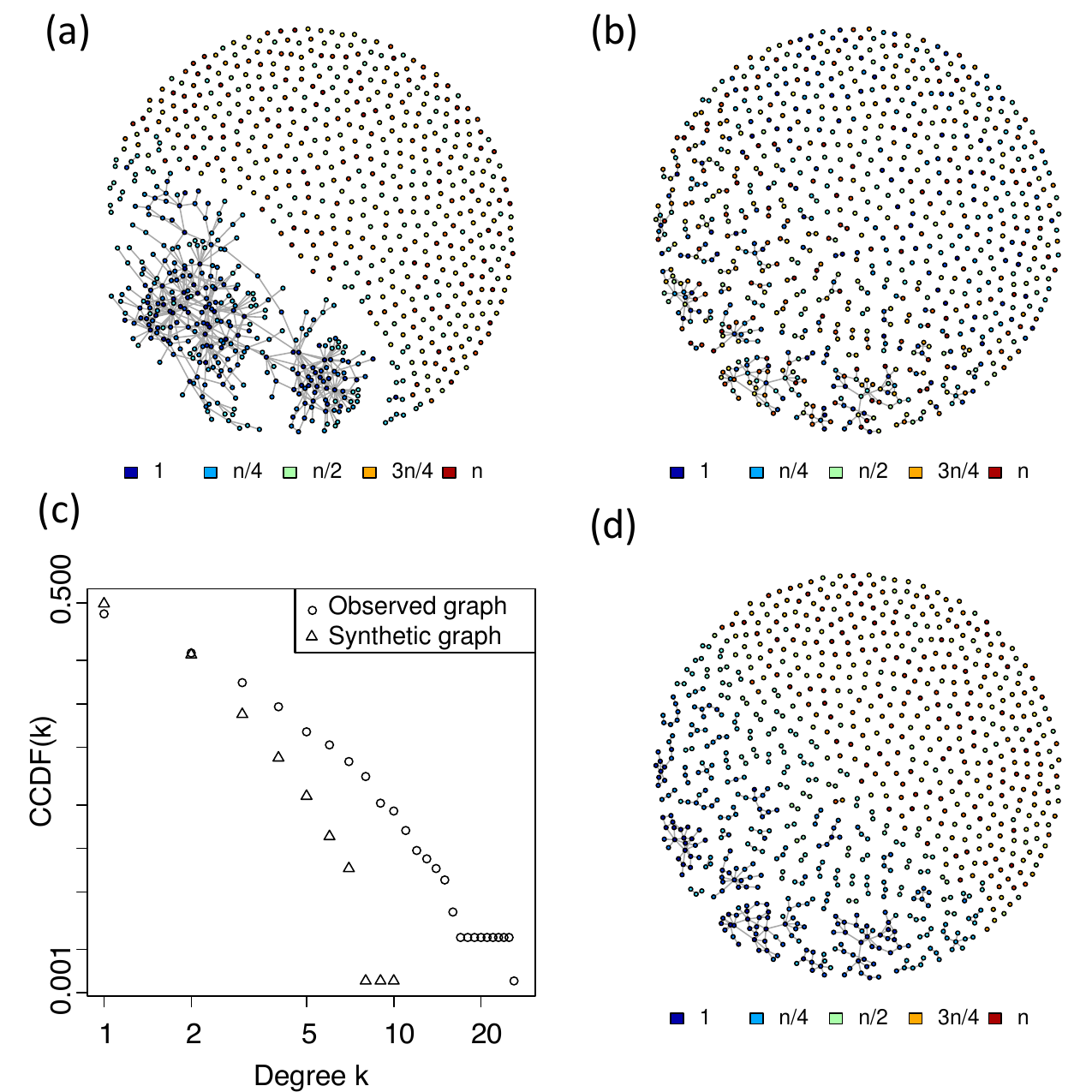}
\caption{\label{figure:synth} (a) The subgraph induced by sampling
10\% of the nodes from the HuRI graph uniformly at random.
Node color corresponds to the estimated order in which the node
entered the graph, with dark blue for the earliest nodes and bright red for the most recent nodes. (b) A DMC graph with the same number of nodes as the graph in (a), and with $q_m$ and $q_c$ equal to the estimated values of $q_m$ and $q_c$ for the graph in (a) ($\hat{q}_m = 0.709$ and $\hat{q}_c = 0.201$). Node color corresponds to the true order in which the node entered the graph. (c) Complementary cumulative distribution function (CCDF) of node degrees  for the graphs in (a) and (b). (d) The same graph as in (b), but with node color corresponding to the estimated order in which the node entered the graph.}
\end{figure}

Results for the HuRI graph and subgraphs are in Table \ref{table:huri}.
Results for the STRING graphs are in Table \ref{table:string}.
The subgraph induced by sampling 10\% of the nodes from HuRI
uniformly at random is in Figure \ref{figure:synth}(a).
There appears to be one large component and many isolated nodes.
The color of each node corresponds to the estimated order in which that node
entered the graph.
The nodes at the center of the large component are estimated to be
the oldest,
and the isolated nodes are estimated to be the newest.
The correlation between the estimated order in which a node entered the graph
and its degree is -0.613 (p $< 10^{-15}$),
meaning early nodes have high degree and recent nodes have low degree.

\begin{table}
\begin{ruledtabular}
\begin{tabular}{cccccc}
\multicolumn{1}{c}{$\bm{p}$} & \multicolumn{1}{c}{\textbf{Nodes}} & \multicolumn{1}{c}{\textbf{Edges}} & \multicolumn{1}{c}{$\bm{\hat{q}_m}$} & \multicolumn{1}{c}{$\bm{\hat{q}_c}$} & \multicolumn{1}{c}{\textbf{Duration (h)}} \\
\hline
0.05 & 414 & 140 & 0.675 & 0.157 & 0.006 \\
0.10 & 827 & 606 & 0.684 & 0.202 & 0.058 \\
0.15 & 1,241 & 1,377 & 0.707 & 0.198 & 0.215 \\
0.20 & 1,654 & 2,263 & 0.706 & 0.214 & 0.545 \\
0.25 & 2,068 & 3,255 & 0.729 & 0.231 & 1.088 \\
0.30 & 2,482 & 4,820 & 0.729 & 0.224 & 2.123 \\
0.35 & 2,895 & 5,930 & 0.749 & 0.228 & 3.855 \\
0.40 & 3,309 & 8,776 & 0.738 & 0.224 & 6.259 \\
0.45 & 3,722 & 10,679 & 0.743 & 0.224 & 9.238 \\
0.50 & 4,136 & 12,581 & 0.755 & 0.213 & 13.203 \\
0.55 & 4,550 & 15,987 & 0.749 & 0.218 & 18.639 \\
0.60 & 4,963 & 18,980 & 0.748 & 0.213 & 25.278 \\
0.65 & 5,377 & 21,340 & 0.751 & 0.224 & 32.951 \\
0.70 & 5,790 & 25,152 & 0.753 & 0.203 & 38.111 \\
0.75 & 6,204 & 30,208 & 0.755 & 0.213 & 52.372 \\
0.80 & 6,618 & 33,631 & 0.755 & 0.209 & 58.366 \\
0.85 & 7,031 & 37,529 & 0.757 & 0.202 & 78.638 \\
0.90 & 7,445 & 41,985 & 0.758 & 0.210 & 94.396 \\
0.95 & 7,858 & 46,604 & 0.759 & 0.200 & 97.277 \\
1.00 & 8,272 & 52,068 & 0.760 & 0.204 & 112.806 \\
\end{tabular}
\end{ruledtabular}
\caption{\label{table:huri}
Results for the HuRI graph and subgraphs.
The Minimize $Y(u,v)$ algorithm was run on only one core, in Python.
Here, $p = \text{proportion of nodes sampled}$.}
\end{table}

\begin{table}
\begin{ruledtabular}
\begin{tabular}{lcccc}
\textbf{Species} & \textbf{Nodes} & \textbf{Edges} & $\bm{\hat{q}_m}$ & $\bm{\hat{q}_c}$ \\
\hline
\textit{C. elegans} & 5,410 & 62,419 & 0.470 & 0.578 \\
\textit{D. melanogaster} & 6,439 & 90,385 & 0.430 & 0.621 \\
\textit{E. coli} & 985 & 4,223 & 0.257 & 0.646 \\
\textit{S. cerevisiae} & 3,557 & 50,198 & 0.341 & 0.708 \\
\end{tabular}
\end{ruledtabular}
\caption{\label{table:string}
Results for STRING graphs.}
\end{table}

The DMC graph generated with $q_m$ and $q_c$ equal to the corresponding
estimates from the graph in Figure \ref{figure:synth}(a)
is in Figure \ref{figure:synth}(b).
There are many medium-sized components and many isolated nodes.
The color of each node corresponds to the true order in which that node
entered the graph.
Old and recent nodes are evenly distributed among the medium-sized components
and isolated nodes.
The correlation between the true order in which a node entered the graph
and its degree is -0.078 (p $= 0.03$),
meaning early nodes have high degree and recent nodes have low degree, although the correlation is weak.

Figure \ref{figure:synth}(c) contains the degree distributions
for the graphs in \ref{figure:synth}(a) and \ref{figure:synth}(b).
The observed graph has more nodes of higher degree.
Figure \ref{figure:synth}(d) is the same as Figure \ref{figure:synth}(b),
except with node color corresponding to the estimated order in which
the node entered the graph.
Here, we use the estimated order used to calculate the Strict Kendall's
$\tau$, i.e., we assume that the algorithm cannot tell which node
in a pair is the anchor node and which is the new node
and thus removes one at random in the deconstruction process.
The Strict Kendall's $\tau$ for this estimated order is 0.016.
The correlation between the estimated order in which a node entered the graph
and its degree is -0.759 (p $< 10^{-15}$),
meaning early nodes have high degree and recent nodes have low degree.
Similar plots, but for the four STRING organisms, are in the Supplemental
Materials.

\section{Discussion}

Mechanistic models are useful representations of the processes
by which networks grow and change over time.
The DMC model, for example,
has proved to be an accurate representation of how protein-protein
interaction networks change over time \cite{navlakha2011,middendorf2005}.
However, the parameters of these models can be difficult to estimate
using maximum likelihood estimation.
This paper proposed a framework for estimating
these parameters and evaluated that framework with the DMC model.

When estimating $q_m$ and $q_c$ in a DMC graph,
one can minimize RMSE and time by using the Minimize $Y(u,v)$ algorithm.
If the goal is to estimate $\theta_n$,
this is the worst algorithm.
No matter which algorithm one chooses,
a naive confidence interval will have low coverage.
More research is needed to determine how to calculate better standard
errors than those one would use for a sequence of independent
Bernoulli experiments.
Averaging across observed values of $\theta_n$ will probably not improve
the estimates of $q_m$ and $q_c$.
The same can be said for using expectation-maximization (EM),
although our results may arise from the fact that we used the maximum
likelihood estimates as initial values for EM,
and did not perturb them.
Using other initial values may yield higher performance for EM.

For graphs with 7 nodes,
the Exhaustive algorithm had the worst RMSE for $q_m$.
This is probably an example of overfitting.
Any time $\hat{q}_m \in \{0,1\}$, the part of the likelihood
containing $\hat{q}_m$ becomes $1$,
and any time $\hat{q}_c \in \{0,1\}$, the part of the likelihood
containing $\hat{q}_c$ becomes $1$.
It is more likely that one or both of $\hat{q}_m$ and $\hat{q}_c$
will be 0 or 1 for some $\theta_n$ in a small graph,
and if such a $\theta_n$ exists,
the Exhaustive algorithm will find it.

It is interesting to note that in Figure \ref{figure:nll},
the estimates of $q_m$ took the shape of an oval,
with most estimates concentrated at the poles.
This was the case even though many of those estimates arose
from the NK algorithm,
which uses initial values in a grid.

Further work can be done to make the algorithms proposed here run faster,
and to find other algorithms with better performance.
But the main question that remains to be answered is whether
the framework proposed here can be applied to other mechanistic
network models.
At first glance, it appears that this framework is restricted to
reversible mechanistic network models.
What happens if, for example, the mechanism involves deleting a node?
Reversing it would require inserting a new node
and connecting it to its former neighbors.
But what if those neighbors cannot be determined from the current state
of the graph?
In this case, our framework can be extended by simulating multiple options,
connecting this new node to neighbors selected through sampling,
and then averaging across these simulations.
Further research is needed to determine the performance of this extension.

\begin{acknowledgments}
The authors would like to thank Alessandro Vespignani, Edoardo Airoldi, and Rui Wang for their helpful suggestions. We would also like to thank John Platig for suggesting the HuRI data. Jonathan Larson was supported by NIH award T32AI007358. Jukka-Pekka Onnela was supported by NIH awards R01AI138901 (Onnela)
and R35CA220523 (Quackenbush).
\end{acknowledgments}


\providecommand{\noopsort}[1]{}\providecommand{\singleletter}[1]{#1}%


\begin{thebibliography}{19}%
\makeatletter
\providecommand \@ifxundefined [1]{%
 \@ifx{#1\undefined}
}%
\providecommand \@ifnum [1]{%
 \ifnum #1\expandafter \@firstoftwo
 \else \expandafter \@secondoftwo
 \fi
}%
\providecommand \@ifx [1]{%
 \ifx #1\expandafter \@firstoftwo
 \else \expandafter \@secondoftwo
 \fi
}%
\providecommand \natexlab [1]{#1}%
\providecommand \enquote  [1]{``#1''}%
\providecommand \bibnamefont  [1]{#1}%
\providecommand \bibfnamefont [1]{#1}%
\providecommand \citenamefont [1]{#1}%
\providecommand \href@noop [0]{\@secondoftwo}%
\providecommand \href [0]{\begingroup \@sanitize@url \@href}%
\providecommand \@href[1]{\@@startlink{#1}\@@href}%
\providecommand \@@href[1]{\endgroup#1\@@endlink}%
\providecommand \@sanitize@url [0]{\catcode `\\12\catcode `\$12\catcode
  `\&12\catcode `\#12\catcode `\^12\catcode `\_12\catcode `\%12\relax}%
\providecommand \@@startlink[1]{}%
\providecommand \@@endlink[0]{}%
\providecommand \url  [0]{\begingroup\@sanitize@url \@url }%
\providecommand \@url [1]{\endgroup\@href {#1}{\urlprefix }}%
\providecommand \urlprefix  [0]{URL }%
\providecommand \Eprint [0]{\href }%
\providecommand \doibase [0]{https://doi.org/}%
\providecommand \selectlanguage [0]{\@gobble}%
\providecommand \bibinfo  [0]{\@secondoftwo}%
\providecommand \bibfield  [0]{\@secondoftwo}%
\providecommand \translation [1]{[#1]}%
\providecommand \BibitemOpen [0]{}%
\providecommand \bibitemStop [0]{}%
\providecommand \bibitemNoStop [0]{.\EOS\space}%
\providecommand \EOS [0]{\spacefactor3000\relax}%
\providecommand \BibitemShut  [1]{\csname bibitem#1\endcsname}%
\let\auto@bib@innerbib\@empty
\bibitem [{\citenamefont {Gilbert}(1959)}]{gilbert1959}%
  \BibitemOpen
  \bibfield  {author} {\bibinfo {author} {\bibfnamefont {E.~N.}\ \bibnamefont
  {Gilbert}},\ }\bibfield  {title} {\bibinfo {title} {Random graphs},\ }\href
  {https://doi.org/10.1214/aoms/1177706098} {\bibfield  {journal} {\bibinfo
  {journal} {The Annals of Mathematical Statistics}\ }\textbf {\bibinfo
  {volume} {30}},\ \bibinfo {pages} {1141 } (\bibinfo {year}
  {1959})}\BibitemShut {NoStop}%
\bibitem [{\citenamefont {V\'{a}zquez}\ \emph {et~al.}(2003)\citenamefont
  {V\'{a}zquez}, \citenamefont {Flammini}, \citenamefont {Maritan},\ and\
  \citenamefont {Vespignani}}]{vazquezetal2003}%
  \BibitemOpen
  \bibfield  {author} {\bibinfo {author} {\bibfnamefont {A.}~\bibnamefont
  {V\'{a}zquez}}, \bibinfo {author} {\bibfnamefont {A.}~\bibnamefont
  {Flammini}}, \bibinfo {author} {\bibfnamefont {A.}~\bibnamefont {Maritan}},\
  and\ \bibinfo {author} {\bibfnamefont {A.}~\bibnamefont {Vespignani}},\
  }\bibfield  {title} {\bibinfo {title} {Modeling of protein interaction
  networks},\ }\href@noop {} {\bibfield  {journal} {\bibinfo  {journal}
  {ComPlexUs}\ }\textbf {\bibinfo {volume} {1}},\ \bibinfo {pages} {38}
  (\bibinfo {year} {2003})}\BibitemShut {NoStop}%
\bibitem [{\citenamefont {V\'azquez}(2003)}]{vazquez2003}%
  \BibitemOpen
  \bibfield  {author} {\bibinfo {author} {\bibfnamefont {A.}~\bibnamefont
  {V\'azquez}},\ }\bibfield  {title} {\bibinfo {title} {Growing network with
  local rules: Preferential attachment, clustering hierarchy, and degree
  correlations},\ }\href {https://doi.org/10.1103/PhysRevE.67.056104}
  {\bibfield  {journal} {\bibinfo  {journal} {Phys. Rev. E}\ }\textbf {\bibinfo
  {volume} {67}},\ \bibinfo {pages} {056104} (\bibinfo {year}
  {2003})}\BibitemShut {NoStop}%
\bibitem [{\citenamefont {Chen}\ and\ \citenamefont {Onnela}(2019)}]{chen2019}%
  \BibitemOpen
  \bibfield  {author} {\bibinfo {author} {\bibfnamefont {S.}~\bibnamefont
  {Chen}}\ and\ \bibinfo {author} {\bibfnamefont {J.-P.}\ \bibnamefont
  {Onnela}},\ }\bibfield  {title} {\bibinfo {title} {A bootstrap method for
  goodness of fit and model selection with a single observed network},\ }\href
  {https://doi.org/10.1038/s41598-019-53166-6} {\bibfield  {journal} {\bibinfo
  {journal} {Scientific Reports}\ }\textbf {\bibinfo {volume} {9}},\ \bibinfo
  {pages} {16674} (\bibinfo {year} {2019})}\BibitemShut {NoStop}%
\bibitem [{\citenamefont {Chen}\ \emph {et~al.}(2020)\citenamefont {Chen},
  \citenamefont {Mira},\ and\ \citenamefont {Onnela}}]{chen2020}%
  \BibitemOpen
  \bibfield  {author} {\bibinfo {author} {\bibfnamefont {S.}~\bibnamefont
  {Chen}}, \bibinfo {author} {\bibfnamefont {A.}~\bibnamefont {Mira}},\ and\
  \bibinfo {author} {\bibfnamefont {J.-P.}\ \bibnamefont {Onnela}},\ }\bibfield
   {title} {\bibinfo {title} {Flexible model selection for mechanistic network
  models},\ }\href {https://doi.org/10.1093/comnet/cnz024} {\bibfield
  {journal} {\bibinfo  {journal} {Journal of Complex Networks}\ }\textbf
  {\bibinfo {volume} {8}},\ \bibinfo {pages} {cnz024} (\bibinfo {year}
  {2020})},\ \Eprint
  {https://arxiv.org/abs/https://academic.oup.com/comnet/article-pdf/8/2/cnz024/33543529/cnz024.pdf}
  {https://academic.oup.com/comnet/article-pdf/8/2/cnz024/33543529/cnz024.pdf}
  \BibitemShut {NoStop}%
\bibitem [{\citenamefont {Raynal}\ \emph {et~al.}(2023)\citenamefont {Raynal},
  \citenamefont {Hoffmann},\ and\ \citenamefont {Onnela}}]{raynal2023}%
  \BibitemOpen
  \bibfield  {author} {\bibinfo {author} {\bibfnamefont {L.}~\bibnamefont
  {Raynal}}, \bibinfo {author} {\bibfnamefont {T.}~\bibnamefont {Hoffmann}},\
  and\ \bibinfo {author} {\bibfnamefont {J.-P.}\ \bibnamefont {Onnela}},\
  }\bibfield  {title} {\bibinfo {title} {Cost-based feature selection for
  network model choice},\ }\href
  {https://doi.org/10.1080/10618600.2022.2151453} {\bibfield  {journal}
  {\bibinfo  {journal} {Journal of Computational and Graphical Statistics}\
  }\textbf {\bibinfo {volume} {0}},\ \bibinfo {pages} {1} (\bibinfo {year}
  {2023})}\BibitemShut {NoStop}%
\bibitem [{\citenamefont {Wiuf}\ \emph {et~al.}(2006)\citenamefont {Wiuf} \emph
  {et~al.}}]{wiuf2006}%
  \BibitemOpen
  \bibfield  {author} {\bibinfo {author} {\bibfnamefont {C.}~\bibnamefont
  {Wiuf}} \emph {et~al.},\ }\bibfield  {title} {\bibinfo {title} {A likelihood
  approach to analysis of network data},\ }\href
  {https://doi.org/10.1073/pnas.0600061103} {\bibfield  {journal} {\bibinfo
  {journal} {Proc. Natl. Acad. Sci.}\ }\textbf {\bibinfo {volume} {103}},\
  \bibinfo {pages} {7566} (\bibinfo {year} {2006})}\BibitemShut {NoStop}%
\bibitem [{\citenamefont {Arnold}\ \emph {et~al.}(2021)\citenamefont {Arnold},
  \citenamefont {Mondrag{\'o}n},\ and\ \citenamefont {Clegg}}]{arnold2021}%
  \BibitemOpen
  \bibfield  {author} {\bibinfo {author} {\bibfnamefont {N.~A.}\ \bibnamefont
  {Arnold}}, \bibinfo {author} {\bibfnamefont {R.~J.}\ \bibnamefont
  {Mondrag{\'o}n}},\ and\ \bibinfo {author} {\bibfnamefont {R.~G.}\
  \bibnamefont {Clegg}},\ }\bibfield  {title} {\bibinfo {title}
  {Likelihood-based approach to discriminate mixtures of network models that
  vary in time},\ }\href@noop {} {\bibfield  {journal} {\bibinfo  {journal}
  {Scientific reports}\ }\textbf {\bibinfo {volume} {11}},\ \bibinfo {pages}
  {1} (\bibinfo {year} {2021})}\BibitemShut {NoStop}%
\bibitem [{\citenamefont {Overgoor}\ \emph {et~al.}(2019)\citenamefont
  {Overgoor}, \citenamefont {Benson},\ and\ \citenamefont
  {Ugander}}]{overgoor2019}%
  \BibitemOpen
  \bibfield  {author} {\bibinfo {author} {\bibfnamefont {J.}~\bibnamefont
  {Overgoor}}, \bibinfo {author} {\bibfnamefont {A.}~\bibnamefont {Benson}},\
  and\ \bibinfo {author} {\bibfnamefont {J.}~\bibnamefont {Ugander}},\
  }\bibfield  {title} {\bibinfo {title} {Choosing to grow a graph: modeling
  network formation as discrete choice},\ }in\ \href@noop {} {\emph {\bibinfo
  {booktitle} {The World Wide Web Conference}}}\ (\bibinfo {year} {2019})\ pp.\
  \bibinfo {pages} {1409--1420}\BibitemShut {NoStop}%
\bibitem [{\citenamefont {Cantwell}\ \emph {et~al.}(2021)\citenamefont
  {Cantwell}, \citenamefont {St-Onge},\ and\ \citenamefont
  {Young}}]{cantwell2021}%
  \BibitemOpen
  \bibfield  {author} {\bibinfo {author} {\bibfnamefont {G.~T.}\ \bibnamefont
  {Cantwell}}, \bibinfo {author} {\bibfnamefont {G.}~\bibnamefont {St-Onge}},\
  and\ \bibinfo {author} {\bibfnamefont {J.-G.}\ \bibnamefont {Young}},\
  }\bibfield  {title} {\bibinfo {title} {Inference, model selection, and the
  combinatorics of growing trees},\ }\href@noop {} {\bibfield  {journal}
  {\bibinfo  {journal} {Physical Review Letters}\ }\textbf {\bibinfo {volume}
  {126}},\ \bibinfo {pages} {038301} (\bibinfo {year} {2021})}\BibitemShut
  {NoStop}%
\bibitem [{\citenamefont {Middendorf}\ \emph {et~al.}(2005)\citenamefont
  {Middendorf}, \citenamefont {Ziv},\ and\ \citenamefont
  {Wiggins}}]{middendorf2005}%
  \BibitemOpen
  \bibfield  {author} {\bibinfo {author} {\bibfnamefont {M.}~\bibnamefont
  {Middendorf}}, \bibinfo {author} {\bibfnamefont {E.}~\bibnamefont {Ziv}},\
  and\ \bibinfo {author} {\bibfnamefont {C.~H.}\ \bibnamefont {Wiggins}},\
  }\bibfield  {title} {\bibinfo {title} {Inferring network mechanisms: The
  drosophila melanogaster protein interaction network},\ }\href
  {https://doi.org/10.1073/pnas.0409515102} {\bibfield  {journal} {\bibinfo
  {journal} {Proceedings of the National Academy of Sciences}\ }\textbf
  {\bibinfo {volume} {102}},\ \bibinfo {pages} {3192} (\bibinfo {year}
  {2005})},\ \Eprint
  {https://arxiv.org/abs/https://www.pnas.org/content/102/9/3192.full.pdf}
  {https://www.pnas.org/content/102/9/3192.full.pdf} \BibitemShut {NoStop}%
\bibitem [{\citenamefont {Navlakha}\ and\ \citenamefont
  {Kingsford}(2011)}]{navlakha2011}%
  \BibitemOpen
  \bibfield  {author} {\bibinfo {author} {\bibfnamefont {S.}~\bibnamefont
  {Navlakha}}\ and\ \bibinfo {author} {\bibfnamefont {C.}~\bibnamefont
  {Kingsford}},\ }\bibfield  {title} {\bibinfo {title} {Network archaeology:
  Uncovering ancient networks from present-day interactions},\ }\href
  {https://doi.org/10.1371/journal.pcbi.1001119} {\bibfield  {journal}
  {\bibinfo  {journal} {PLOS Computational Biology}\ }\textbf {\bibinfo
  {volume} {7}},\ \bibinfo {pages} {1} (\bibinfo {year} {2011})}\BibitemShut
  {NoStop}%
\bibitem [{\citenamefont {Dempster}\ \emph {et~al.}(1977)\citenamefont
  {Dempster}, \citenamefont {Laird},\ and\ \citenamefont
  {Rubin}}]{dempster1977}%
  \BibitemOpen
  \bibfield  {author} {\bibinfo {author} {\bibfnamefont {A.~P.}\ \bibnamefont
  {Dempster}}, \bibinfo {author} {\bibfnamefont {N.~M.}\ \bibnamefont
  {Laird}},\ and\ \bibinfo {author} {\bibfnamefont {D.~B.}\ \bibnamefont
  {Rubin}},\ }\bibfield  {title} {\bibinfo {title} {Maximum likelihood from
  incomplete data via the em algorithm},\ }\href@noop {} {\bibfield  {journal}
  {\bibinfo  {journal} {Journal of the Royal Statistical Society Series B
  (Methodological)}\ }\textbf {\bibinfo {volume} {39}},\ \bibinfo {pages} {1}
  (\bibinfo {year} {1977})}\BibitemShut {NoStop}%
\bibitem [{\citenamefont {Meng}(2009)}]{meng2009}%
  \BibitemOpen
  \bibfield  {author} {\bibinfo {author} {\bibfnamefont {X.-L.}\ \bibnamefont
  {Meng}},\ }\bibfield  {title} {\bibinfo {title} {{Decoding the
  H-likelihood}},\ }\href {https://doi.org/10.1214/09-STS277C} {\bibfield
  {journal} {\bibinfo  {journal} {Statistical Science}\ }\textbf {\bibinfo
  {volume} {24}},\ \bibinfo {pages} {280 } (\bibinfo {year}
  {2009})}\BibitemShut {NoStop}%
\bibitem [{\citenamefont {Kendall}(1938)}]{kendall1938}%
  \BibitemOpen
  \bibfield  {author} {\bibinfo {author} {\bibfnamefont {M.~G.}\ \bibnamefont
  {Kendall}},\ }\bibfield  {title} {\bibinfo {title} {A new measure of rank
  correlation},\ }\href {https://doi.org/10.1093/biomet/30.1-2.81} {\bibfield
  {journal} {\bibinfo  {journal} {Biometrika}\ }\textbf {\bibinfo {volume}
  {30}},\ \bibinfo {pages} {81} (\bibinfo {year} {1938})},\ \Eprint
  {https://arxiv.org/abs/https://academic.oup.com/biomet/article-pdf/30/1-2/81/423380/30-1-2-81.pdf}
  {https://academic.oup.com/biomet/article-pdf/30/1-2/81/423380/30-1-2-81.pdf}
  \BibitemShut {NoStop}%
\bibitem [{\citenamefont {Luck~\textit{et al.}}(2020)}]{luck2020}%
  \BibitemOpen
  \bibfield  {author} {\bibinfo {author} {\bibfnamefont {K.}~\bibnamefont
  {Luck~\textit{et al.}}},\ }\bibfield  {title} {\bibinfo {title} {A reference
  map of the human binary protein interactome},\ }\href@noop {} {\bibfield
  {journal} {\bibinfo  {journal} {Nature}\ }\textbf {\bibinfo {volume} {580}},\
  \bibinfo {pages} {402} (\bibinfo {year} {2020})}\BibitemShut {NoStop}%
\bibitem [{\citenamefont {DFCI{\ }Center{\ }for{\ }Cancer{\ }Systems{\
  }Biology}(2021)}]{interactome2021}%
  \BibitemOpen
  \bibfield  {author} {\bibinfo {author} {\bibnamefont {DFCI{\ }Center{\ }for{\
  }Cancer{\ }Systems{\ }Biology}},\ }\href {http://www.interactome-atlas.org/}
  {\bibinfo {title} {The human reference protein interactome mapping project}}
  (\bibinfo {year} {2021}),\ \bibinfo {note}
  {\url{http://www.interactome-atlas.org/}}\BibitemShut {NoStop}%
\bibitem [{\citenamefont {Szklarczyk~\textit{et al.}}(2018)}]{szklarczyk2018}%
  \BibitemOpen
  \bibfield  {author} {\bibinfo {author} {\bibfnamefont {D.}~\bibnamefont
  {Szklarczyk~\textit{et al.}}},\ }\bibfield  {title} {\bibinfo {title}
  {{STRING v11: protein?protein association networks with increased coverage,
  supporting functional discovery in genome-wide experimental datasets}},\
  }\href {https://doi.org/10.1093/nar/gky1131} {\bibfield  {journal} {\bibinfo
  {journal} {Nucleic Acids Research}\ }\textbf {\bibinfo {volume} {47}},\
  \bibinfo {pages} {D607} (\bibinfo {year} {2018})},\ \Eprint
  {https://arxiv.org/abs/https://academic.oup.com/nar/article-pdf/47/D1/D607/27437323/gky1131.pdf}
  {https://academic.oup.com/nar/article-pdf/47/D1/D607/27437323/gky1131.pdf}
  \BibitemShut {NoStop}%
\bibitem [{\citenamefont {{R Core Team}}(2019)}]{r2019}%
  \BibitemOpen
  \bibfield  {author} {\bibinfo {author} {\bibnamefont {{R Core Team}}},\
  }\href {https://www.R-project.org/} {\emph {\bibinfo {title} {R: A Language
  and Environment for Statistical Computing}}},\ \bibinfo {organization} {R
  Foundation for Statistical Computing},\ \bibinfo {address} {Vienna, Austria}
  (\bibinfo {year} {2019})\BibitemShut {NoStop}%
\end{thebibliography}
\end{document}